\documentclass[
aps,prd,
12pt,
nopreprintnumbers,
showpacs,
eqsecnum,
nofootinbib
]{revtex4-1}

\usepackage{graphicx}
\usepackage{amssymb}

\begin{document}

\title{Discrete heat kernel, UV modif{}ied Green's function, and higher derivative
theories}
\author{Nahomi Kan}\email[]{kan@gifu-nct.ac.jp}
\affiliation{National Institute of Technology, Gifu College,
Motosu-shi, Gifu 501-0495, Japan
}
\author{Masashi Kuniyasu}\email[]{mkuni13@yamaguchi-u.ac.jp}
\author{Kiyoshi Shiraishi}\email[]{shiraish@yamaguchi-u.ac.jp}
\author{Zhenyuan Wu}\email[]{b501wb@yamaguchi-u.ac.jp}
\affiliation{
Graduate School of Sciences and Technology for Innovation, Yamaguchi
University, Yamaguchi-shi, Yamaguchi 753--8512, Japan}
\date{\today}

\begin{abstract}
We perform the UV deformation of the Green's function in free scalar field theory
using a discrete heat kernel method. It is found that the simplest UV deformation
based on the discretized diffusion equation leads to the well-known
Pauli--Villars effective Lagrangian. Furthermore, by extending assumptions on the
discretized equation, we find that the general higher derivative theory is derived
from the present UV deformation. In some specific cases, we also calculate the
vacuum expectation values of the scalar field squared in nontrivial background
spaces and examine their dependence on the UV cutoff constant.
\end{abstract}



\maketitle

\section{Introduction}
\label{introduction}
Taking the effects of quantum gravity into account, we can expect the emergence of
fundamental lengths in effective field theories. In recent years, various types of
UV deformed Green's functions incorporating a fundamental length have been proposed
based on the expression of the Green's function obtained by the heat kernel method
\cite{Pad1,Pad2,Pad3,Pad4,Pad5,Pad6,SSP,KSSP,AD,ABM,AL,Siegel}.%
\footnote{The authors of
Refs.~\cite{Pad1,Pad2,Pad3,Pad4,Pad5,Pad6,SSP,KSSP,AD,ABM,AL} explored the
construction of them mainly assuming the duality in the path integral amplitude.}
Putting a lower limit of integral on the Schwinger's proper-time parameter in the
standard expression is known as the easiest way to regularize the Green's function
in the heat kernel method. Siegel
\cite{Siegel} pointed out that this treatment is equivalent to considering
infinite derivative effective field theory instead of original canonical theory.
 
When we consider UV deformation from the viewpoint of regularization, we can
imagine that there can be too many variations in the way of deformation. However,
it is desirable to maintain properties related to the symmetry of the original
field theory as much as possible.
 
In our previous paper \cite{KKSW}, we calculated the vacuum expectation values of
the stress tensor in nontrivial background spaces using the UV deformed Green's
function \cite{Pad1,Pad2,Pad3,Pad4,Pad5,Pad6,SSP,KSSP,AD,ABM,AL,Siegel}. The
dependence on the cutoff constant (which may be identified with the Planck length)
was clarified by using numerical calculations. However, a primitive approach, in
which we applied only the raw deformation of scalar Green's functions to the
original theory, resulted in small violations of the conservation law, which is
negligible in the limit of a small cutoff length and large physical scales.
 
In the present paper, we start with modifying the original heat kernel method. We
consider a heat kernel that satisfies a discretized heat diffusion equation. Of
course, we first define the basic scheme which leads to the standard 
Green's function without change. Then, by changing the lower
limit of the sum, instead of integral, UV deformation of the Green's
function is realized. In this case, the Green's function containing higher order
terms in the momentum space is obtained. It coincides with
Pauli--Villars' Green function in its simplest form \cite{IZ,Collins,PS}.%
\footnote{The Pauli--Villars-type scalar field model is also considered \cite{SS}
to study the possible effects of high energy corrections to the inflationary
perturbation spectrum.} We identify the higher derivative theory as the effective
field theory that yields the UV modified Green's function. 

In the second half of the paper, based on the formulation used in the first half,
we consider more general UV deformation and evaluate the vacuum expectation values
of the scalar field squared in nontrivial background spaces as specific cases. In
each case, attention is paid to the cutoff dependence of the physical
quantity. The dimensionful cutoff constant can be taken as a
parameter in a mere regularization, but similarly to the introduction of the
aforementioned UV deformed Green's function
\cite{Pad1,Pad2,Pad3,Pad4,Pad5,Pad6,SSP,KSSP,AD,ABM,AL,Siegel}, it may be resulted
from yet-unknown quantum gravity theory (possibly directly related to the Planck
length). Thus, the cutoff constant can be considered as a fundamental constant in
nature. The discussion in this paper is the first stage of `phenomenological'
demonstration which is aimed to future theories with absolute UV completion.
 
The structure of this paper is as follows. In Sec.~\ref{sec2}, we describe the
Green's function constructed from the heat kernel which satisfies the discretized
equations. In this paper, we consider a locally flat Euclidean space as a
background space, and here we consider a uniformly isotropic $D$-dimensional space
$\mathbf{R}^D$. Then, UV deformation is introduced based on a simple assumption. 
We discuss more generalized UV modification in Sec.~\ref{sec3}. In the
effective theory due to the UV modification obtained in the previous sections, we
calculate the vacuum expectation values of the square of the scalar field in two
kinds of nontrivial background spaces, a Kaluza--Klein-type space and a conical
space in Sec.~\ref{sec4}. In the last section, we discuss the results and future
prospects. 

\section{Discrete heat kernel and the simplest UV modified Green's function}
\label{sec2}

First of all, we review some basic facts of the standard Green's functions
and heat kernels \cite{Vassilevich,Camporesi}.
The standard Green's function
 $G(x,x')$ of a scalar
field with mass $m$ in a metric space with the metric $g_{\mu\nu}$ satisfies
\begin{equation}
(-\Box_x+m^2){G}(x,x')=\frac{1}{\sqrt{g}}\delta(x,x')\,,
\end{equation}
where $\Box=\frac{1}{\sqrt{g}}\partial_\mu\sqrt{g}g^{\mu\nu}\partial_\nu$ is 
the d'Alembert operator on scalar fields, and
$\frac{1}{\sqrt{g}}\delta(x,x')$ is the $D$-dimensional covariant delta function.

The main tool that we select here is the heat kernel of the Laplacian.
The heat kernel $K(x,x';s)$ is introduced as a representation of the
Green's function constructed by using it:
\begin{equation}
{G}(x,x')=\int_0^\infty ds\, K(x,x';s)\,.
\end{equation}
Here, the heat kernel follows the equation
\begin{equation}
\left[\frac{\partial}{\partial s}+(-\Box_x+m^2)\right]K(x,x';s)=0\,,
\end{equation}
with the initial condition $\lim_{s\rightarrow
0}K(x,x';s)=\frac{1}{\sqrt{g}}\delta(x,x')$. 

Next, we consider the Fourier transform $\tilde{K}(p;s)$ of the heat kernel in
an Euclidean space $\mathbf{R}^D$. This is defined through the relation
\begin{equation}
K(x,x';s)=K(x-x';s)=\int\frac{d^Dp}{(2\pi)^D} \tilde{K}(p;s) e^{ip\cdot
(x-x')}\,.
\end{equation}
Then, the heat equation becomes
\begin{equation}
\left[\frac{\partial}{\partial s}+(p^2+m^2)\right]\tilde{K}(p;s)=0\,,
\end{equation}
with the `initial' condition
$\tilde{K}(p;0)=1$.
One can easily find the solution
\begin{equation}
\tilde{K}(p;s)=e^{-s(p^2+m^2)}\,,
\end{equation}
and immediately obtain the Fourier transform of the Green's function
$\tilde{G}(p)$ in
$\mathbf{R}^D$ as follows:
\begin{equation}
\tilde{G}(p)=\int_0^\infty \exp\left[-s (p^2+m^2)\right] ds
=\frac{1}{p^2+m^2}\,.
\label{integ}
\end{equation}
Then, the real space Green's function is found to be
\begin{equation}
G(x,x')=G(x-x')=\int\frac{d^Dp}{(2\pi)^D} e^{ip\cdot
(x-x')}\tilde{G}(p)=\int\frac{d^Dp}{(2\pi)^D} \frac{e^{ip\cdot
(x-x')} }{p^2+m^2}\,.
\end{equation}
We can exchange the order of integrations. That is, the real space heat kernel is
\begin{equation}
K(x-x';s)=\frac{1}{(4\pi s)^{D/2}}\exp\left({-\frac{|x-x'|^2}{4s}-m^2 s}\right)\,,
\end{equation}
and the Green's function is recovered as
\begin{equation}
G(x-x')=\int_0^\infty ds\, K(x-x';s)=\int_0^\infty ds\frac{1}{(4\pi
s)^{D/2}}\exp\left({-\frac{|x-x'|^2}{4s}-m^2 s}\right)\,.
\end{equation}
Anyway, the closed form for $G(x-x')$ turns out to be
\begin{equation}
G(x-x')=\frac{m^{D-2}}{(2\pi)^{D/2}(m|x-x'|)^{\frac{D-2}{2}}}
K_{\frac{D-2}{2}}({ m|x-x'|})\,,
\end{equation}
where $K_\nu(z)$ is the modified Bessel function of the second kind.
Incidentally, for the massless case ($m=0$), we find
\begin{equation}
G(x-x')=\frac{\Gamma\left({\textstyle
\frac{D-2}{2}}\right)}{4\pi^{D/2}|x-x'|^{D-2}}\,,
\label{normm0}
\end{equation}
where $\Gamma(z)$ is the gamma function.


Now, we turn to considering a discretized heat kernel formulation,
whose continuous limit reproduces the standard method constructing the Green's
function. To this end, we propose the discrete proper-time heat equation in
$\mathbf{R}^D$ as%
\footnote{Here we used the backward difference, but using the forward difference
does not give essentially different results.}
\begin{equation}
\frac{\tilde{K}_{k}-\tilde{K}_{k-1}}{\epsilon}=
-(p^2+m^2)\tilde{K}_{k}\,,
\label{rr}
\end{equation}
where $k=1,\dots,N$ and $\epsilon$ is a positive constant.
We adopt the initial value $\tilde{K}_{0}=1$.
Then, the solution of the recurrence equation (\ref{rr}) is given by
\begin{equation}
\tilde{K}_{k}=\left[\frac{1}{1+
\epsilon(p^2+m^2)}\right]^k
\,.
\end{equation}
On the other hand, the following sum is directly concluded from (\ref{rr}):
\begin{equation}
(p^2+m^2)\left[
\sum_{k=1}^{N}\epsilon\tilde{K}_{k}\right]=
\tilde{K}_{0}-\tilde{K}_{N}
\,.
\end{equation}
Observing that $\lim_{N\rightarrow\infty}\left[{1+
\epsilon(p^2+m^2)}\right]^{-N}=0$, we find
\begin{equation}
\sum_{k=1}^\infty\epsilon\tilde{K}_k=
\frac{1}{p^2+m^2}
\,,
\end{equation}
which is just the discretized version of the integral (\ref{integ}).

The continuum limit is realized by taking
$k\epsilon\rightarrow s$, $\tilde{K}_k\rightarrow \tilde{K}(p;s)$, and
$\epsilon\rightarrow 0$. Then, we find that
\begin{equation}
\left[\frac{1}{1+
\epsilon(p^2+m^2)}\right]^k\rightarrow\exp[-s(p^2+m^2)]
\,,\quad
\sum_{k=1}^\infty\epsilon\tilde{K}_k\rightarrow
\int_0^\infty \tilde{K}(p;s) ds=\frac{1}{p^2+m^2}
\,.
\end{equation}

After making the above preparation, we consider the UV modification of the Green's
function. For this purpose, we define the modified Green's function in momentum
space
\begin{equation}
\tilde{G}_n(p)\equiv\sum_{k=n}^\infty\epsilon\tilde{K}_k
=\frac{1}{(p^2+m^2)[1+\epsilon(p^2+m^2)]^{n-1}}
\,.
\end{equation}
To see the effect of the modification, we investigate a special case, in which
we assume that $\epsilon=\epsilon'/n$ and $n\rightarrow\infty$. In this case, 
$\tilde{G}$ reduces to
\begin{equation}
\tilde{G}_\infty(p)
=\frac{e^{-\epsilon'(p^2+m^2)}}{p^2+m^2}
=\int_{\epsilon'}^\infty \exp[-(p^2+m^2)s] ds 
\,,
\end{equation}
which was studied by Siegel \cite{Siegel}. It is also known that the present manner
is used in the UV regularization in the standard heat kernel formalism.

Here, we study the simplest modification, the case with $n=2$. Namely, we use
\begin{equation}
\tilde{G}_2(p)
=\frac{1}{(p^2+m^2)[1+\epsilon(p^2+m^2)]}
=\frac{1}{p^2+m^2}-\frac{1}{p^2+m^2+1/\epsilon}
\,.
\end{equation}
This is the simplest Pauli--Villars-type propagator (Green's function)
\cite{IZ,Collins,PS}. The effective scalar field theory that leads to this
Green's function is
\begin{equation}
S_2=\int\frac{d^Dp}{(2\pi)^D}\,\tilde{\cal
L}_2=-\frac{1}{2}\int\frac{d^Dp}{(2\pi)^D}\tilde{\phi}(-p)(p^2+m^2)
[1+\epsilon(p^2+m^2)]
\tilde{\phi}(p)
\,,
\end{equation}
in the Fourier transformed form, and in the real coordinate space, it turns out to
be
\begin{equation}
S_2=\int d^Dx \,{\cal
L}_2=-\frac{1}{2}\int d^Dx \,{\phi}(x)(-\Box+m^2)
[1+\epsilon(-\Box+m^2)]
{\phi}(x)
\,,
\end{equation}
where $\phi$ is the real scalar field and $\tilde{\phi}$ is its Fourier transform.

We can introduce an auxiliary field to continue study,
referring 
to 
works on higher derivative theories \cite{GOW,CL} motivated by the
seminal Lee--Wick model
\cite{LW}.
We consider the alternative Lagrangian
\begin{equation}
{\cal
L}'_2=-\frac{1}{2}\phi(-\Box+m^2)\phi-
\psi(-\Box+m^2)\phi+\frac{1}{2\epsilon}\psi^2
\,.
\end{equation}
The equation of motion of the initially auxiliary field $\psi$ from the Lagrangian
is
\begin{equation}
\psi=\epsilon(-\Box+m^2)\phi
\,.
\end{equation}
Thus, substituting this equation to ${\cal L}'_2$ yields the original Lagrangian
${\cal L}_2$. On the other hand, setting $\chi\equiv\phi+\psi$, we find
\begin{equation}
{\cal
L}'_2=-\frac{1}{2}\chi(-\Box+m^2)\chi+\frac{1}{2}\psi(-\Box+m^2+1/\epsilon)\psi
\,.
\end{equation}
This effective Lagrangian describes two free scalar fields, whose masses are
$m$ and $\sqrt{m^2+1/\epsilon}$. The scalar field $\psi$ with mass
$\sqrt{m^2+1/\epsilon}$ governed by the part of the Lagrangian with a `wrong'
sign. This field decouples from the physical spectrum if $\epsilon\rightarrow 0$,
since its mass becomes infinitely large. 


The real space Green's function 
\begin{equation}
G_2(x-x')=\frac{m^{\frac{D-2}{2}}K_{\frac{D-2}{2}}({
m|x-x'|})-(m^2+1/\epsilon)^{\frac{D-2}{4}}K_{\frac{D-2}{2}}({
\sqrt{m^2+1/\epsilon}|x-x'|})}{(2\pi)^{D/2}|x-x'|^{\frac{D-2}{2}}}
\,,
\end{equation}
thus involves two mass scales $m$ and $\sqrt{m^2+1/\epsilon}$.
For the massless case ($m=0$), we find
\begin{equation}
G_2(x-x')=\frac{\Gamma\left({\textstyle
\frac{D-2}{2}}\right)}{4\pi^{D/2}|x-x'|^{D-2}}-\frac{\epsilon^{-\frac{D-2}{4}}K_{\frac{D-2}{2}}({
\epsilon^{-\frac{1}{2}}|x-x'|})]}{(2\pi)^{D/2}|x-x'|^{\frac{D-2}{2}}}\,.
\end{equation}
At small $|x-x'|$, the Green's function $G_2(x-x')$ behaves as
$\sim\epsilon^{-1}|x-x'|^{4-D}$ for $D\ne 4$ and
$\sim\epsilon^{-1}\ln(\epsilon^{-1/2}|x-x'|)$
for $D=4$.


In the next section, we will consider possible generalization of the UV
modification along with the present line of thought.

\section{generalization of discretization and interesting cases}
\label{sec3}

The generalization of (\ref{rr}) is 
not difficult.
The parameter $\epsilon$
can have dependence on the integer $k$, i.e.,
\begin{equation}
\frac{\tilde{K}_{k}-\tilde{K}_{k-1}}{\epsilon_{k}}=
-(p^2+m^2)\tilde{K}_{k}\,,
\label{rr2}
\end{equation}
where positive parameters $\epsilon_k$ are restricted by
\begin{equation}
\epsilon_k=\epsilon_n\quad\mbox{for~} k>n\,.
\label{rr3}
\end{equation}
Then, the summation yields 
\begin{equation}
\tilde{G}_n(p)\equiv\sum_{k=n}^\infty\epsilon_k\tilde{K}_k
=\frac{1}{(p^2+m^2)\prod_{k=1}^{n-1}[1+\epsilon_k(p^2+m^2)]}
\,.
\end{equation}
Note that $\tilde{G}_1(p)
=\frac{1}{p^2+m^2}$ and $\tilde{G}_2(p)
=\frac{1}{(p^2+m^2)[1+\epsilon_1(p^2+m^2)]}$. They are the same as those discussed
in Sec.~\ref{sec2}. Now, $\tilde{G}_n(p)$ involves $n-1$ parameters in general.


The effective action that leads to this modified propagator in an expression with
the momentum argument is
\begin{equation}
S=-\frac{1}{2}\int\frac{d^Dp}{(2\pi)^D}\tilde{\phi}(-p)[\tilde{G}_n(p)]^{-1}
\tilde{\phi}(p)
\,,
\label{eft}
\end{equation}
since the free propagator is obtained by the Gaussian integration in path integral
formalism.

Provided that values of $\epsilon_k$ are all
different, the Green's function in the momentum space can be expanded by partial
fractions as follows:
\begin{equation}
\frac{1}{(p^2+m^2)\prod_{k=1}^{n-1}[1+\epsilon_k(p^2+m^2)]}
=\frac{1}{p^2+m^2}+\sum_{k=1}^{n-1}\frac{c_k\epsilon_k}{1+\epsilon_k(p^2+m^2)}
\,,
\label{pfe}
\end{equation}
where the coefficient $c_k$ reads
\begin{equation}
c_k=\frac{-\epsilon_k^{n-2}}{\prod_{i\ne k}^{n-1}(\epsilon_k-\epsilon_i)}
\,.
\end{equation}


Now, we consider concrete examples. 
We here consider two particular cases for $\{\epsilon_k\}$.
We shall call them model A and model B.

\smallskip
\noindent
$\bullet$ Model A

First, we consider $\{\epsilon_k\}=\{\epsilon,
\frac{\epsilon}{2},\frac{\epsilon}{3},
\ldots
\frac{\epsilon}{n-1}\}$.
We know the mathematical formula
\begin{equation}
\tilde{G}_A(p)\equiv\frac{1}{(p^2+m^2)\prod_{k=1}^{n-1}[1+\frac{\epsilon}{k}(p^2+m^2)]}
=\sum_{k=0}^{n-1}{\textstyle
{n-1}\choose{k}}\frac{(-1)^k\epsilon/k}{1+\frac{\epsilon}{k}(p^2+m^2)}
\,,
\end{equation}
thus, 
\begin{equation}
c_k=(-1)^k{\textstyle
{n-1}\choose{k}}\qquad(k=1,\ldots,n-1)\,.
\end{equation}
Note that the case $n=2$ trivially reduces to the simplest case treated in
Sec.~\ref{sec2}. Thus, this first example is a straightforward extension of the
simplest case.

\smallskip
\noindent
$\bullet$ Model B

The second example enjoys the infinite derivatives, i.e.,
defined with $n\rightarrow\infty$.
We set
$\{\epsilon_k\}=\{\epsilon,
\frac{\epsilon}{2^2},\frac{\epsilon}{3^2},
\ldots\}$. Now, the Green's function in the momentum space becomes
\begin{eqnarray}
\tilde{G}_B(p)\equiv\frac{1}{(p^2+m^2)\prod_{k=1}^{\infty}[1+\frac{\epsilon}{k^2}(p^2+m^2)]}
&=&\frac{\pi\sqrt{\epsilon}}{\sqrt{p^2+m^2}\sinh[\pi\sqrt{\epsilon(p^2+m^2)}]}
\nonumber \\
&=&\frac{1}{p^2+m^2}+\sum_{k=1}^{\infty}
\frac{2(-1)^k\epsilon/k^2}{1+\frac{\epsilon}{k^2}(p^2+m^2)}
\,,
\end{eqnarray}
and we find
\begin{equation}
c_k=(-1)^k \cdot 2\qquad (k\ge 1)\,.
\end{equation}
This type of the Green's function has appeared in Refs.~\cite{FS1,FS2,Fujikawa}
as generalizations of the Pauli--Villars regularization. Incidentally,
the Green's function for Model B with $m=0$ in configuration space can be written
as
\begin{equation}
G_B(x-x')=\frac{\sqrt{\epsilon}}{\pi^{\frac{D-1}{2}}}\sum_{k=0}^\infty
\frac{\Gamma\left(\frac{D-1}{2}\right)}{\left[|x-x'|^2+4\pi^2\left(k+\frac{1}{2}
\right)^2\epsilon\right]^{\frac{D-1}{2}}}\qquad(\mbox{for~} m=0)\,.
\end{equation}
The limit $\epsilon\rightarrow 0$ reduces this expression to the normal
massless Green's function (\ref{normm0}).
Furthermore, especially for $D=5$, one can obtain a closed form of $G_B(w)$ as
\begin{equation}
G_B(w)=\frac{1}{8\pi^{2}}
\left[\frac{1}{w^3}\tanh\frac{w}{2l}-\frac{1}{2lw^2}\mbox{sech}^2\frac{w}{2l}
\right]\qquad(\mbox{for~} m=0, D=5)\,,
\end{equation}
where $w\equiv|x-x'|$.

In the next section, we evaluate vacuum expectation values of the scalar field
squared in the effective theory of Models A and B in nontrivial spaces.

\section{vacuum expectation value of the scalar field squared for the UV
modified cases in nonsimply connected spaces}
\label{sec4}

Now, we can express the one-loop vacuum
expectation value of the square of the scalar field $\langle\phi^2\rangle$, or
also called as the quantum fluctuation of the scalar field, in the effective
higher derivative theory. We find, with the aid of (\ref{pfe}),
\begin{equation}
\langle\phi^2\rangle=\lim_{x'\rightarrow x}G_n(x,x')=\lim_{x'\rightarrow
x}\left[G(x,x';m^2)+\sum_{k=1}^{n-1} c_k G(x,x';M^2_k)\right]\,,
\end{equation}
where $G(x,x';M^2)$ is the Green's function of a canonical scalar field with
mass $M$ and $M^2_k=m^2+1/\epsilon_k$.

We should note that the value of $\langle \phi^2\rangle$ is still a
`bare' quantity. Even if the highest divergence (e.g. quadratic divergences in
four dimensions) has been canceled, divergences remain for a sufficient
high dimensional case (see the calculation in this section below). Moreover, even
if divergences have ceased, the bare value of the stress tensor should be of order
of
$\epsilon^{-\alpha}$ $(\alpha>0)$. If one would
like to combine the cutoff constant
$\epsilon$ with some fundamental theory, the finite but huge quantities that
appears in quantum expectation values should be subtracted by some method.
In the present analysis, similar to many past works including our work \cite{KKSW},
subtraction is performed by hand, but done on the basis of physical meaning.

The calculation of the vacuum expectation values of
the free scalar field squared in nonsimply connected space has been studied by many
authors in recent decades. We consider two specific cases. We pick up a
Kaluza--Klein-type space and a conical space.

\subsection{The Kaluza--Klein-type space}

First, we consider a $D$-dimensional space whose metric is given by
\begin{equation}
ds^2=dx^\lambda dx_\lambda=\sum_{i=1}^{D-1}dz_i^2+dy^2\,,
\end{equation}
where the Greek index runs over $1,\dots,D$, while the Roman index runs over
$1,\dots,D-1$. The coordinate
$y$ is the coordinate on a circle ($S^1$) and we assume that $0\le y<L$, where $L$
is the circumference of the circle. The periodic boundary condition in the
direction of
$y$ is supposed to be applied on the massless scalar, i.e., $\phi(y+L)=\phi(y)$.
In this space, $\mathbf{R}^{D-1}\otimes S^1$, the standard massless Green's
function for a scalar field takes the form
\begin{equation}
G(x,x';M^2)=\int_0^\infty\frac{ds}{(4\pi
s)^{D/2}}\sum_{q=-\infty}^{\infty}\exp\left[\textstyle
-\frac{\zeta^2+(\eta-qL)^2}{4s}-M^2s\right]
\,,
\end{equation}
where $\zeta=|z-z'|$ and $\eta=y-y'$.
Then, we define the quantities with subtraction
\begin{equation}
\langle
\bar{\phi^2}\rangle[M^2]\equiv\lim_{x'\rightarrow
x}
\bar{G}(x,x';M^2)\,,
\label{subtt}
\end{equation}
where
$\bar{G}(x,x';M^2)\equiv{G}(x,x';M^2)-\lim_{L\rightarrow\infty}{G}(x,x';M^2)$.
This subtraction is aimed to take the difference from the quantity in the
infinitely extending space $\mathbf{R}^D$.
After straightforward manipulations, we get
\begin{equation}
\langle
\bar{\phi^2}\rangle[M^2]=\frac{2}{(4\pi)^{D/2}}\int_0^\infty\frac{ds}{
s^{D/2}}\sum_{q=1}^{\infty}\exp\left[{\textstyle -\frac{q^2L^2}{4s}-M^2s}\right]
=\frac{1}{\pi}\sum_{q=1}^\infty\left(\textstyle \frac{M}{2\pi
qL}\right)^{\frac{D-2}{2}} K_{\frac{D-2}{2}}(qML)
\,,
\end{equation}
and thus, we can evaluate the vacuum expectation value in the effective theory as
\begin{equation}
\langle\bar{\phi^2}\rangle=\langle
\bar{\phi^2}\rangle[m^2]+\sum_{k=1}^{n-1} c_k\langle
\bar{\phi^2}\rangle[M_k^2]\,.
\label{sfs}
\end{equation}

In general, in the higher derivative scalar field theory (\ref{eft}) which leads
to the Green's function (\ref{pfe}), the vacuum expectation value for the scalar
field squared
$\langle\bar{\phi^2}\rangle$ is written by
\begin{equation}
\langle
\bar{\phi^2}\rangle=\frac{2}{(4\pi)^{D/2}}\int_0^\infty\frac{ds}{
s^{D/2}}\sum_{q=1}^{\infty}\exp\left[{\textstyle -\frac{q^2L^2}{4s}}-m^2 s\right]
\varrho(s)
\,,
\end{equation}
where $\varrho(s)\equiv 1+\sum_{k=1}^{n-1}c_k\exp[-s/\epsilon_k]$.

\noindent
For Model A, we find
\begin{equation}
\varrho_A(s)=\sum_{k=0}^{n-1}(-1)^k{\textstyle
{n-1}\choose{k}}e^{-ks/\epsilon}=(1-e^{-s/\epsilon})^{n-1}
\,.
\end{equation} 
For Model B, we find
\begin{equation}
\varrho_B(s)=\sum_{k=-\infty}^\infty (-1)^k
e^{-k^2s/\epsilon}=\vartheta_4(0,e^{-s/\epsilon})
=\sqrt{\frac{\pi\epsilon}{s}}\vartheta_2(0,e^{-\pi^2\epsilon/s})
=\sqrt{\frac{\pi\epsilon}{s}}\sum_{k=-\infty}^\infty
e^{-\pi^2(k+1/2)^2\epsilon/s}
\,,
\end{equation}
where $\vartheta_j(v,q)$ is the Jacobi theta function.


We now numerically evaluate $\langle
\bar{\phi^2}\rangle$ for the case with $m=0$ and $D=5$.
We define a cutoff length scale, $l\equiv\sqrt{\epsilon}$
and use it hereafter.

\begin{figure}[ht]
\centering
\includegraphics
{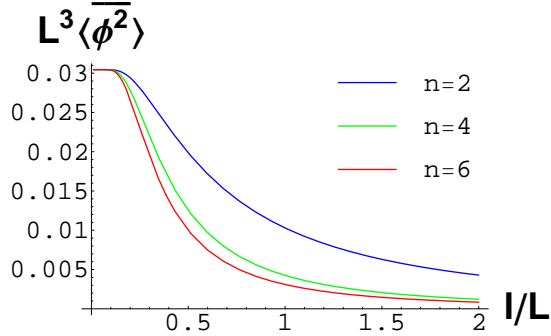}
\caption{The vacuum expectation values of the square of the scalar field
regulated as
$L^3\langle\bar{\phi^2}\rangle$ for
$m=0$ and $D=5$ are plotted against the cutoff length $l$ divided by the
circumference
$L$ of
$S^1$ in the cases with $n=2$, $n=4$, and $n=6$ in Model A.}
\label{fig1}
\end{figure}
\begin{figure}[ht]
\centering
\includegraphics
{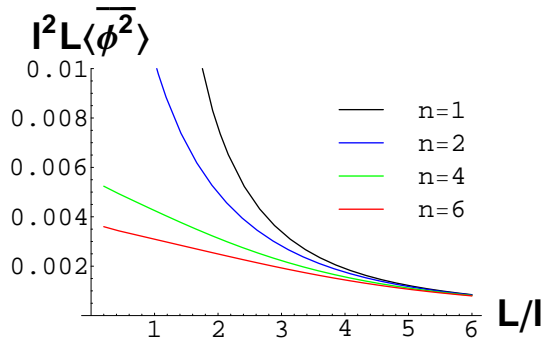}
\caption{The vacuum expectation values of the square of the scalar field
$l^2L\langle\bar{\phi^2}\rangle$ for $m=0$ and $D=5$ are plotted against the
circumference
$L$ of $S^1$ divided by the cutoff length
$l$ in the cases with $n=1$, $n=2$, $n=4$, and $n=6$ in Model A. The black curve
for the $n=1$ case indicates the usual case, i.e., the case that the cutoff length
is set to zero. }
\label{fig2}
\end{figure}

Fig.~\ref{fig1} shows $L^3\langle\bar{\phi^2}\rangle$ in Model A as a function
of $l/L$ for
$m=0$ and $D=5$. Note that $\lim_{l\rightarrow 0}L^3\langle\bar{\phi^2}\rangle=
\frac{\zeta_R(D-2)}{2\pi^{D/2}}$, where $\zeta_R(z)$ is Riemann's zeta
function, and it is
$\approx 0.0343574$ for
$D=5$. One can see that
$L^3\langle\bar{\phi^2}\rangle$ monotonically decreases as $l$ increases.
Fig.~\ref{fig2} shows
$l^2L\langle\bar{\phi^2}\rangle$ in Model A as a function of
$L/l$ for
$m=0$ and $D=5$. 
The value of $l^{D-3}L\langle\bar{\phi^2}\rangle$ in $D$ dimensions at $L=0$
becomes finite if
$n>\frac{D-1}{2}$, and is expressed by the following integral as for Model A:
\begin{equation}
\lim_{L/l\rightarrow 0}l^{D-3}L\langle\bar{\phi^2}\rangle=
\frac{1}{(4\pi)^{D-1}}\int_0^\infty\frac{(1-e^{-t})^{n-1}}{t^{\frac{D-1}{2}}}dt\,.
\end{equation}
For $D=5$ and $n=4$, $\lim_{L/l\rightarrow 0}l^{2}L\langle\bar{\phi^2}\rangle_4
\approx 0.0054653$, while for $D=5$ and $n=6$, $\lim_{L/l\rightarrow
0}l^{2}L\langle\bar{\phi^2}\rangle_6
\approx 0.00369362$.

\begin{figure}[ht]
\centering
\includegraphics
{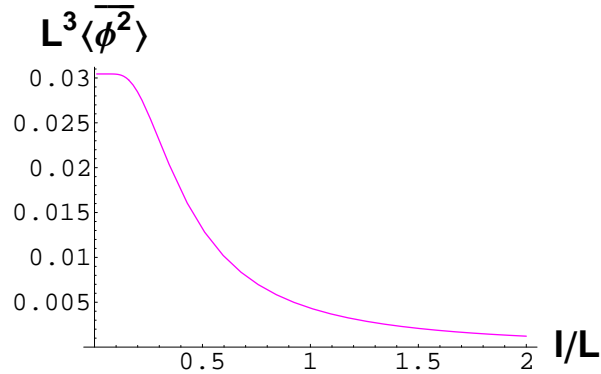}
\caption{The vacuum expectation value regulated as $L^2\langle\bar{\phi^2}\rangle$
for $m=0$ and $D=5$ are plotted against the cutoff length $l$ divided by the
circumference $L$ of $S^1$ in Model B.}
\label{fig3}
\end{figure}
\begin{figure}[ht]
\centering
\includegraphics
{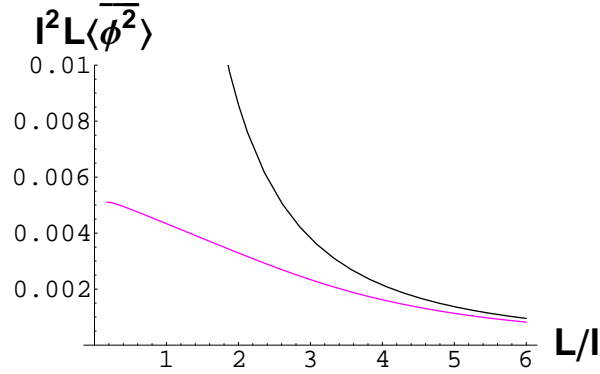}
\caption{The vacuum expectation value regulated as $L^2\langle\bar{\phi^2}\rangle$
for $m=0$ and $D=5$ are plotted 
against the circumference $L$ of $S^1$ divided by the cutoff length $l$ in Model B.
The black curve indicates the canonical case, i.e., the case that
the cutoff length is set to zero and the other curve indicate the present UV
modified case of Model B.}
\label{fig4}
\end{figure}

For Model B, we find the expression of the vacuum expectation value
$\langle\bar{\phi^2}\rangle=\bar{G}_B(0)$ in $\mathbf{R}^{D-1}\otimes S^1$ for
$m=0$:
\begin{equation}
\langle\bar{\phi^2}\rangle=\frac{2l}{\pi^{\frac{D-1}{2}}}\sum_{q=1}^\infty
\sum_{k=0}^\infty
\frac{\Gamma\left(\frac{D-1}{2}\right)}{\left[q^2L^2+4\pi^2\left(k+\frac{1}{2}
\right)^2l^2\right]^{\frac{D-1}{2}}}\,.
\end{equation}
In Fig.~\ref{fig3}, the value of $L^3\langle\bar{\phi^2}\rangle$ in Model B with
$D=5$ and $m=0$ is plotted against
$l/L$. Fig.~\ref{fig4} shows $l^2L\langle\bar{\phi^2}\rangle$ in Model B with
$D=5$ and $m=0$ 
against $L/l$. The limit $L\rightarrow 0$ gives a finite value for
$l^{D-3}L\langle\bar{\phi^2}\rangle$ in Model B, in arbitrary dimensions. It turns
out to be
\begin{equation}
\lim_{L/l\rightarrow
0}l^{D-3}L\langle\bar{\phi^2}\rangle=\frac{1-2^{-(D-2)}}{\pi^{D-2}}
\frac{\Gamma(\frac{D-2}{2})\zeta_R(D-2)}{\pi^{\frac{D-2}{2}}}\,.
\end{equation}
In particular, we find
$\lim_{L/l\rightarrow
0}l^{2}L\langle\bar{\phi^2}\rangle=\frac{7\zeta_R(3)}{16\pi^4}\approx
0.00539888$.


\subsection{The conical space}

Next, we consider the vacuum expectation values in a conical space.
A conical space, or a space with a conical singularity at the coordinate origin,
is described by the metric
\begin{equation}
ds^2=\sum_{i=1}^{D-2}(dz^i)^2+dr^2+\frac{r^2}{\nu^2}d\theta^2\,,
\end{equation}
where $\nu$ is a constant greater than unity. This metric is equivalent to
\begin{equation}
ds^2=\sum_{i=1}^{D-2}(dz^i)^2+dr^2+r^2 d\tilde{\theta}^2\,,
\label{flat}
\end{equation}
where the range of $\tilde{\theta}$ is $0<\tilde{\theta}\le 2\pi/\nu$.
This metric adequately describes a locally flat Euclidean space except for
the coordinate origin if $\nu\ne 1$. 

The standard Green's function in a conical space (without a fundamental length) is
known
\cite{Smith,SH,CKV,Moreira}. 
Because it is natural to take $\bar{G}(x,x';M^2)=G(x,x';M^2)-G(x,x';M^2)|_{\nu=1}$,
we find \cite{KKSW}
\begin{eqnarray}
& &\bar{G}(x,x';M^2)=\bar{G}(r,r',\varphi,\zeta;M^2)\nonumber \\
&\equiv&\int_0^\infty ds\, {\textstyle 
\frac{e^{-\frac{r^2+{r'}^2+\zeta^2}{4s}-M^2s
}}{2\pi(4\pi
s)^{D/2}}}\int_{0}^\infty
e^{-\frac{rr'}{2s}\cosh v}\left[{\textstyle
\frac{\nu\sin\nu(\tilde{\varphi}-\pi)}{\cosh\nu
v-\cos\nu(\tilde{\varphi}-\pi)}-\frac{\nu\sin\nu(\tilde{\varphi}+\pi)}{\cosh\nu
v-\cos\nu(\tilde{\varphi}+\pi)}}\right]dv\,,
\label{sk}
\end{eqnarray}
where $x=(r,\theta, z^i)$, $x'=(r',\theta', z'^i)$, and
$\tilde{\varphi}=\tilde{\theta}-\tilde{\theta}'=\varphi/\nu=(\theta-\theta')/\nu$.
Moreover, the vacuum expectation value of a single, canonical 
free scalar field squared can be rewritten as
$\langle
\bar{\phi^2}\rangle[M^2]=\bar{G}(x,x;M^2)$.
Using this, the vacuum expectation value is reduced to be in a similar form
to (\ref{sfs}) and equivalently
\begin{equation}
\langle
\bar{\phi^2}\rangle=\sum_{k=0}^{n-1} c_k\bar{G}(x,x;M_k^2)\,,
\end{equation}
where we define $c_0\equiv 1$ and $M_0\equiv m$, in addition.

In the present case of the conical space, we thus find
\begin{equation}
\langle
\bar{\phi^2}\rangle=\int_0^\infty ds\,\varrho(s) {
\frac{e^{-\frac{r^2}{2s}-m^2 s
}}{2\pi(4\pi
s)^{D/2}}}\int_{0}^\infty
e^{-\frac{r^2}{2s}\cosh v}\left[{
\frac{-2\nu\sin\nu\pi}{\cosh\nu
v-\cos\nu\pi}}\right]dv\,,
\label{ef2}
\end{equation}
where we again used $\varrho(s)\equiv\sum_{k=0}^{n-1}c_k\exp[-s/\epsilon_k]$, which
is already defined in the previous subsection.

Now, we can evaluate the vacuum expectation values of $\langle
\bar{\phi^2}\rangle$ in Models A
and B, very efficiently using these expressions (\ref{ef2}).
We assume that the original, unmodified theory is a free massless scalar theory
($m=0$).

\begin{figure}[ht]
\centering
\includegraphics[width=5cm]{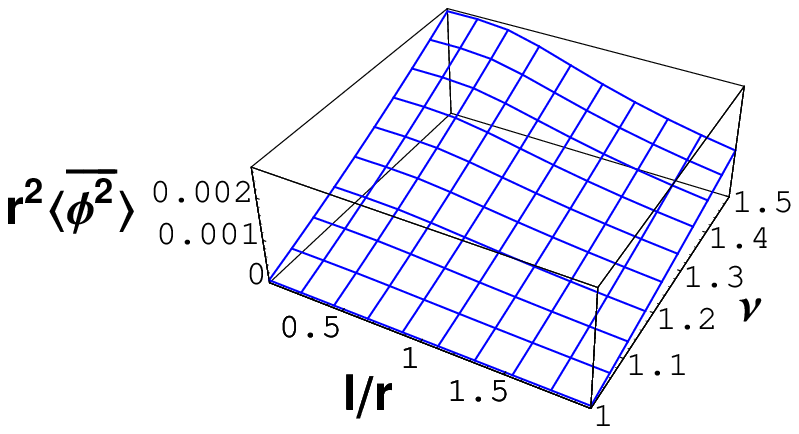}
\includegraphics[width=5cm]{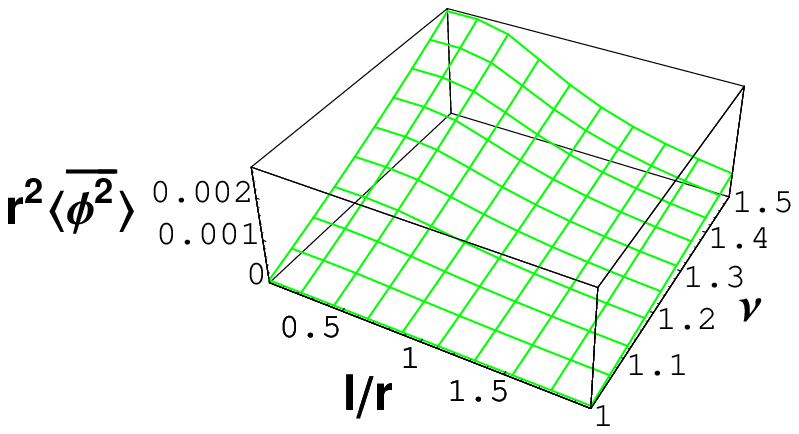}
\includegraphics[width=5cm]{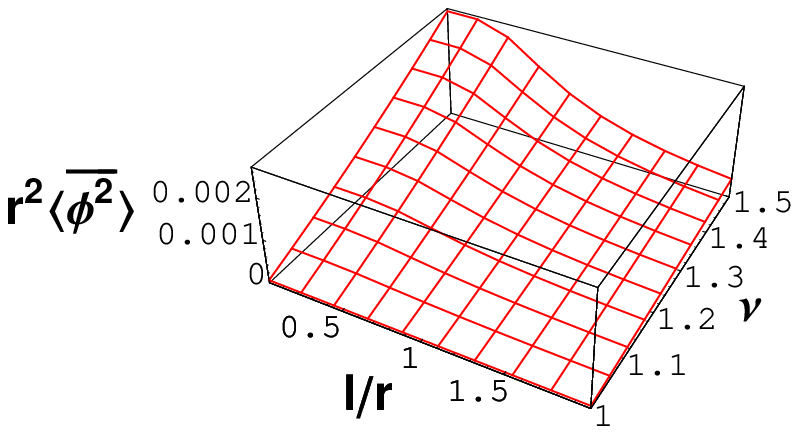}\\
\hspace{1cm} (a) \hspace{4.5cm} (b) \hspace{4.5cm} (c)
 \caption{The vacuum expectation values regulated as $r^2\langle
\bar{\phi^2}\rangle$ for
$m=0$ and $D=4$ are plotted against the cutoff length $l$ divided by the distance
from the origin $r$ in Model A, where (a) for $n=2$,
(b) for $n=4$, and (c) for
$n=6$.}
\label{fig5}
\end{figure}
\begin{figure}[ht]
\centering
\includegraphics[width=5cm]{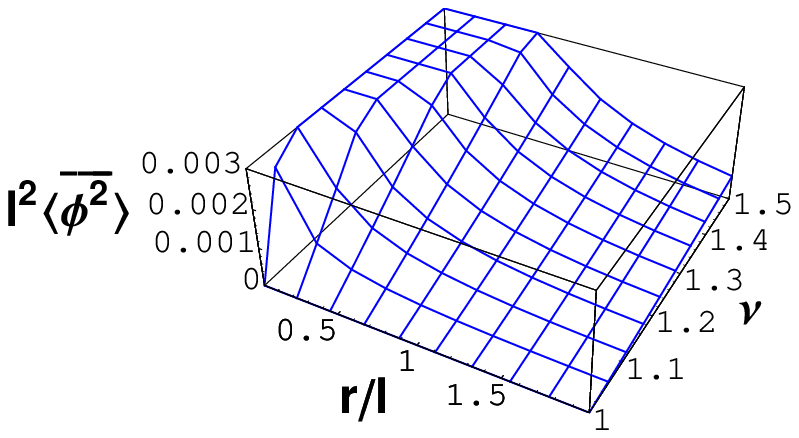}
\includegraphics[width=5cm]{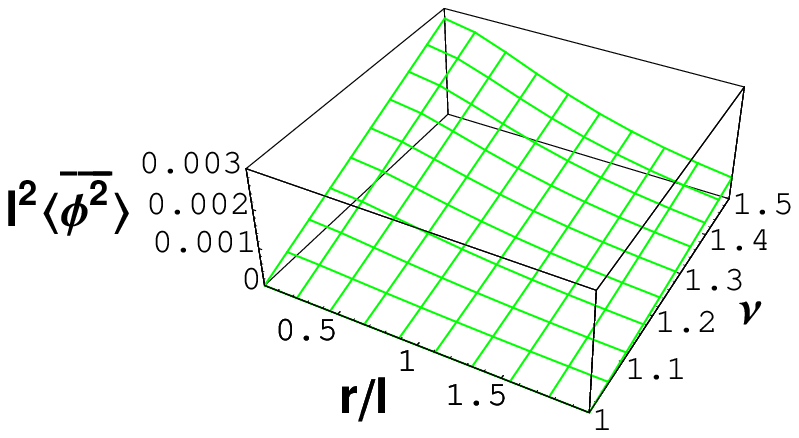}
\includegraphics[width=5cm]{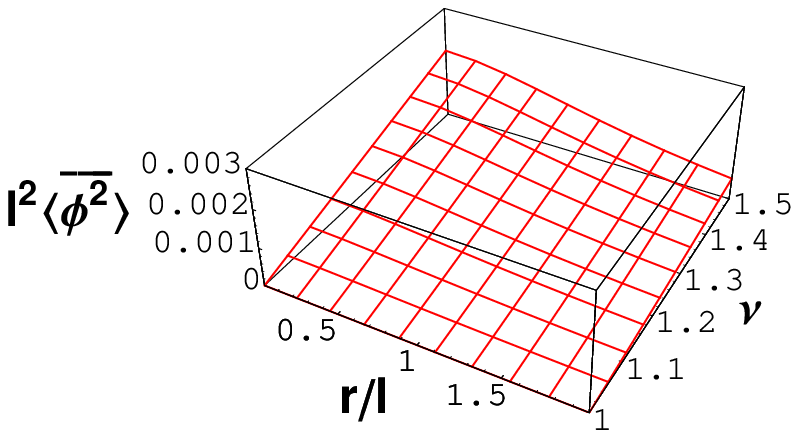}\\
\hspace{1cm} (a) \hspace{4.5cm} (b) \hspace{4.5cm} (c)
 \caption{The vacuum expectation values regulated as $l^2\langle
\bar{\phi^2}\rangle$ for
$m=0$ and $D=4$ are plotted against 
the distance from the origin $r$ divided by the cutoff length $l$ in
Model A, where (a) for
$n=2$, (b) for $n=4$, and (c) for
$n=6$. Note that the values diverge near $r/l\sim 0$ for $n=2$,
the numerical values near $r/l$ are expressed up to a finite cut in the figure
(a).}
\label{fig6}
\end{figure}

In Fig.~\ref{fig5}, we illustrate the $l/r$ dependence of the vacuum expectation
values of the square of the scalar field in Model A. The correction due to
the scale $l$ has a little dependence on
$n$ for a small $l/r$. Note that $\lim_{l/r\rightarrow 0}r^2\langle
\bar{\phi^2}\rangle=\frac{\nu^2-1}{48\pi^2}$ \cite{Smith}. The vacuum expectation
values become small faster for larger $n$, though the changes are qualitatively
the same.

In Figs.~\ref{fig6}, we illustrate the $r/l$ dependence of the vacuum expectation
values in Model A.
In general, when $n>D/2$, the vacuum expectation values are finite at
$r=0$.%
\footnote{Note that, in Fig.~\ref{fig6} (a), the values near $r=0$ is cut by a
certain finite value.}
The value at the origin is finite as
\begin{equation}
\lim_{r/l\rightarrow 0}l^{D-2}\langle
\bar{\phi^2}\rangle=\frac{(\nu-1)l^{D-2}}{(4\pi)^{D/2}}
\int_0^\infty\frac{ds}{s^{D/2}}\varrho_A(s)
=\frac{\nu-1}{(4\pi)^{D/2}}
\int_0^\infty\frac{ds}{t^{D/2}}(1-e^{-t})^{n-1}\,.
\end{equation}
For $D=4$ and $n=4$, this becomes $\approx 0.0054653\times (\nu-1)$,
and for $D=4$ and $n=6$, this becomes $\approx 0.00369362\times (\nu-1)$.

\begin{figure}[ht]
\centering
\includegraphics[width=5cm]{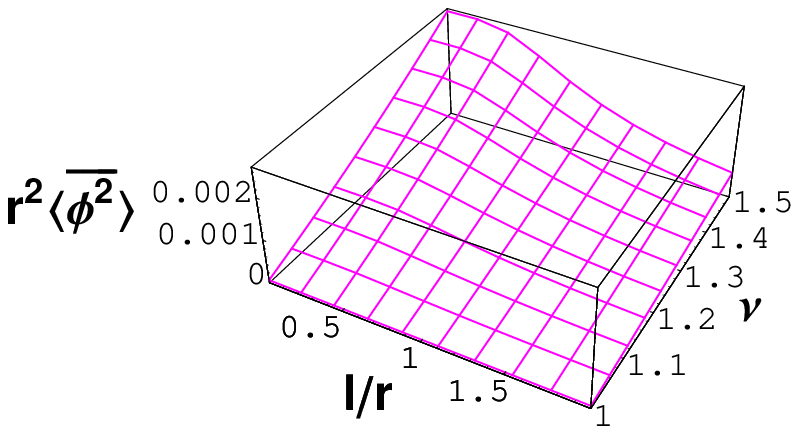}
\includegraphics[width=5cm]{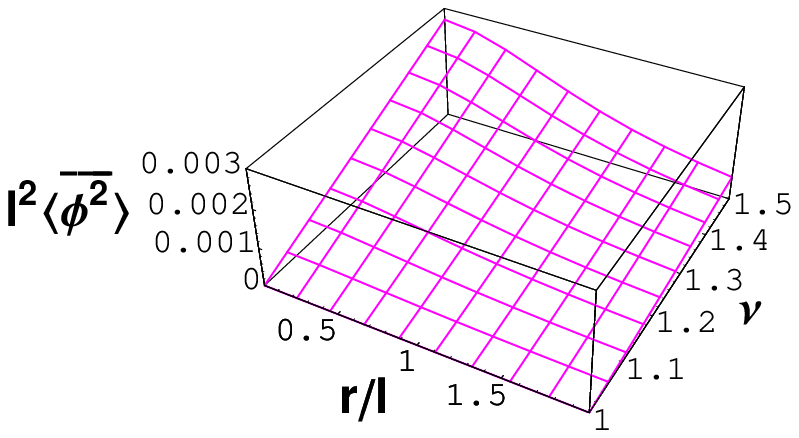}\\
\hspace{1cm} (I) \hspace{4.5cm} (II) 
 \caption{(I) The vacuum expectation values for
$m=0$ and $D=4$ regulated as $r^2\langle
\bar{\phi^2}\rangle$ are plotted against the cutoff length $l$ divided by 
the distance from the
origin $r$. (II) The vacuum expectation values regulated as $l^2\langle
\bar{\phi^2}\rangle$ for $m=0$ and $D=4$ are plotted against 
the distance from the origin $r$ divided by the cutoff length $l$ in Model B.}
\label{fig7}
\end{figure}

In Fig.~\ref{fig7}, we illustrate the $l/r$ dependence and the $r/l$ dependence of
the vacuum expectation values in Model B. The values at $r=0$ are finite.
In both cases, qualitative features are almost the same as in the case of Model A.
The value at the origin is finite as
\begin{equation}
\lim_{r/l\rightarrow 0}l^{D-2}\langle
\bar{\phi^2}\rangle=\frac{\nu-1}{(4\pi)^{D/2}}\int_0^\infty\frac{ds}{s^{D/2}}\varrho_B(s)
=(\nu-1)\frac{1-2^{-(D-1)}}{\pi^{D-1}}
\frac{\Gamma(\frac{D-1}{2})\zeta_R(D-1)}{\pi^{\frac{D-1}{2}}}\,.
\end{equation}
For $D=4$, this becomes $\approx 0.00539888\times (\nu-1)$.

\section{Conclusion}
\label{conclusion}

In this paper, we first introduced a discrete heat kernel formulation of the
Green's function for a free scalar field, and considered the UV deformation that
can be naturally obtained from it. The effective Lagrangian that derives this
UV modified Green's function has been found to be higher derivative theory. In the
simplest case, it corresponds to a simple Pauli--Villars model, or, a Lee--Wick
scalar field theory. In a further generalization, it was found that the present UV
deformation yields a theory involving arbitrarily higher order derivatives, and
the vacuum expectation values of the scalar field squared were
evaluated in nontrivial spaces. We demonstrated that the cutoff length has an
influence on the physical quantities at a short distance in the configuration
space.

The advantage of using the heat kernel method is that we can keep the symmetry of
the space (or break it at will). It is also easy to incorporate the effect of a
constant external field (including spatial metric). On the other hand, one of the
disadvantages is the handling of general interactions of dynamical fields. In
addition, although the form of UV deformation is limited in our approach, it is
also a problem that the sequence $\epsilon_k$ introduced in this paper has a wide
range of arbitrariness other than the case of the two models.

To eliminate UV divergences completely, higher order derivatives that satisfy the
conditions depending on the number of dimensions are required. Even if there is no
divergence, a constant depending on the cutoff constant comes out.
Therefore, some mechanism for canceling the finite contribution in the UV
limit is still required. However, the present demonstration shows that the finite
cutoff dependence of physical quantities occurs when the typical scale in the
considered system is very close to the cutoff scale.
 
In the effective higher derivative theory, a scalar field with a negative norm
appears, so the addition of interaction may bring about instability and violate
unitarity in the Lorentzian spacetime. This is a subject that has been frequently
addressed and has recently been actively discussed \cite{AP,Anselmi}, and in
particular, infinite derivative theory is being studied in various situations
\cite{Modesto,BT,BLM,BLMTY}. One of the models we have proposed (Model B) involves
infinite derivatives, and the unitarity of the theory and its generalization should
be studied further.
Unfortunately, our current understanding has not led us to discuss the unitarity
of the models presented in this paper in more detail.
We consider that detailed discussions about nonlinear terms, i.e., interaction
terms of fields are needed in the (infinitely) higher derivative theory, because
it seems important to properly allocate higher order derivatives to the part of
propagation and the part of interaction vertices.
 
It is interesting to study whether the theory with a basic cutoff scale can be
derived naturally, for example from a model of quantum gravity, or not.%
\footnote{Some specific models were discussed in the last decade
\cite{WZ1,WZ2,RS}.} That is a
topic for the future, but one way might be to proceed with the tentative goal of
deriving behavior like the simple models we examined here.

\bibliographystyle{apsrev4-1}

\end{document}